\documentclass[pra]{revtex4}
\usepackage[dvips]{graphicx}
\usepackage{graphicx,epsfig}
\usepackage{amsmath,amsfonts,amssymb,amsthm,amsbsy,mathtools,mathrsfs}
\usepackage{bbold}

\newcommand{\ket}[1]{\mathop{\left|#1\right>}\nolimits}            
\newcommand{\bra}[1]{\mathop{\left<#1\,\right|}\nolimits}         

\newcommand{\brk}[2]{\langle #1 | #2 \rangle}
\newcommand{\kbr}[2]{| #1\rangle\!\langle #2 |}

\newcommand{\id}{\operatorname{id}}
\newcommand{\oxb}{\overset{\sim}{\otimes}}                

\newcommand{\Tr}[1]{\mathop{{\mathrm{Tr}}_{#1}}}          
\newcommand{\diag}{\mathop{{\mathrm{diag}}}}

\def\bbC{\mathbb{C}}
\def\bbZ{\mathbb{Z}}

\def\FF{\mathscr{F}}

\def\dg{\dagger}

\def\vr{\varrho}
\def\vp{\varphi}

\def\B{\mathcal{B}}

\def\N{\mathcal{N}}

\begin{document}

\author{Kamil Br\'adler}
\email{kbradler@cs.mcgill.ca}
\affiliation{
    School of Computer Science,
    McGill University,
    Montreal, Quebec, H3A 2A7, Canada
    }

\title{On two misconceptions in current relativistic quantum information}

\begin{abstract}
We describe two problems current relativistic quantum information suffers from. The first point is an explanation of an alleged ambiguity of entropic quantities detected in a number of publications and incorrectly resolved in~\cite{aunruh-fermi}. We found that the problem arises due to wrong algebraic manipulations with fermions and ignoring the superselection rule for bosons and fermions. This leads to a misinterpretation of certain entropic quantities when applied to fermion fields. The second discussed point is to alert to a conceptual misunderstanding of the role of entanglement (and quantum correlations in general) in some of the studied relativistic scenarios. Instead, we argue in favor of investigating capacities of quantum channels induced by the relevant physical processes as dictated by quantum Shannon theory.
\end{abstract}

\maketitle

In the first point we identify and clarify a serious error when dealing with the algebra of fermions. According to~\cite{privatecom} the same problem arises in a number of articles dealing with fermion systems in non-inertial frames~\cite{aunruh-fermi,fermi1,fermi1a,fermi2}. We will show that the situation is more complex since there are more than one culprits behind the different results found in the literature.

The relativistic scenario investigated in the relevant articles~\cite{fermi1,fermi1a,fermi2} is similar to the following setup. A Minkowski observer  prepares a fermion multi-particle  state and a Rindler observer (uniformly accelerating) analyzes how the prepared state has changed due to Unruh effect. In particular, one asks about the amount of classical or quantum correlations present and what the dependence on the Rindler observer's proper acceleration is. The correlations are usually analyzed for the three main parties: Minkowski and Rindler observers and a hypothetical observer beyond the horizon. One can investigate any bipartite correlations or even tripartite correlations as has been done in abundance~\cite{fermi1,fermi1a,fermi2}. Due to the local similarity between Rindler and Schwarzschild spacetime~\cite{salas} closely related conclusions can be also drawn for the black hole case. The role of the acceleration parameter is played by the black hole surface gravity.

The problem found in~\cite{aunruh-fermi} always appears in the context of Unruh effect for fermion fields. In fact it is not necessary to consider any relativistic scenario at all. The main issue is the presence of fermion states and how they are manipulated. The Unruh scenario just happened to be an unlucky stage set.

In the second discussed point we argue that for quantum communication purposes the calculations of entanglement measures is misleading and should be abandoned in favor of some more advanced tools. Here the relativistic scenario described above is relevant since the main motivation is naturally  transmission of quantum information.


This article contains no or only a little original research. It is more an `essay' informing about certain issues found in current relativistic quantum information. In this text we deal with various subjects from several different areas of (mathematical) physics: quantum field theory (QFT), quantum groups and quantum Shannon theory. The goal of the text is not to explain all the present concepts and definitions. The details can be found in~\cite{QFT,salas,unruh,braiding,shannon} and references within.

\section{Mistreated fermions}

The authors of~\cite{aunruh-fermi} made an ominous observation. It was pointed out that certain entropic quantities designed to quantify the amount of quantum correlations of multi-partite quantum states do not agree with one another~\cite{fermi1,fermi1a,fermi2}. An explanation of the apparent disparity in the analysis of quantum correlations  for systems of fermions in~\cite{fermi1,fermi1a,fermi2}  was offered in~\cite{aunruh-fermi}. We want to show that the explanation is incorrect and investigate the real cause of the problem.

Calculations of quantum correlations for fermion systems is known to be a subtle issue and has been satisfactorily dealt with in the past~\cite{fermions}. There are three main ingredients one has to be aware of:
\begin{enumerate}
  \item Non-trivial tensor product structure for distinguishable fermions and the concept of braiding.
  \item[2a.] The superselection rule for bosons and fermions.
  \item[2b.] Interpretation of entropic quantities for fermions.
\end{enumerate}
There are several great expositions on the first point~\cite{braiding} and many QFT textbooks discussing the second point~\cite{QFT}. We will see that the authors of~\cite{aunruh-fermi,fermi1} largely ignore the first two points and they can be immediately refuted based on a close inspection motivated by the upcoming discussion. But contrary to their claim~\cite{privatecom}, the rest of the literature~\cite{fermi1a,fermi2} seems to be aware of the first two points (at least those non-overlapping with the authors of~\cite{aunruh-fermi,fermi1}). So why is there so many different conclusions in~\cite{fermi1a,fermi2}? The problem appears to be not in the QFT part of~\cite{fermi1a,fermi2} but rather on the quantum information side. We cannot recalculate all the results presented there but as we discuss further, the problem has likely arisen due to the failure to see the link between the superselection rule for bosons and fermions and the interpretation of certain entropic quantities applied to fermions. To illustrate the problem we will later take a closer look at~\cite{fermi1a}.

We start by presenting the minimal amount of information needed to study systems of fermions. Note that in this text we strictly use the Fock formalism and so there is no doubt whether a fermion state is mode-entangled or not~\cite{indistin} (note that the issues discussed here are not related to the problem of entanglement of indistinguishable particles).  Also, when we talk about particles or modes we mean the same thing. We are allowed to do so since every mode contains at most one particle (fermion).

Fermion annihilation and creation operators ($a,a^\dg$) satisfy the canonical anticommutation relations that are also known as the CAR algebra
\begin{equation}\label{eq:CAR}
    [a,a^\dg]_+=1,
\end{equation}
where $[,]_+$ stands for the anticommutator. Different modes (fermions) will be labeled by different letters ($a,b$ and $c$) and they hide the two relevant fermion degrees of freedom (the momentum and spin). The intermode relations read
\begin{equation}\label{eq:interCAR}
    [a,b]_+=[a,b^\dg]_+=[a^\dg,b^\dg]_+=0
\end{equation}
and similarly for the $c$ mode. Therefore, when we act on a vacuum with $a^\dg b^\dg$, the order of the operators matters
\begin{equation}\label{eq:order}
    a^\dg b^\dg\ket{vac}=\ket{11}_{ab}\neq\ket{11}_{ba}.
\end{equation}
We conclude that
\begin{equation}\label{eq:notensor}
    \ket{11}_{ab}\neq\ket{1}_a\otimes\ket{1}_b,
\end{equation}
where $\otimes$ denotes the standard tensor product. This is an obvious statement -- only in the case of two qubits or two distinguishable bosons the order is irrelevant and $\ket{11}_{ab}=\ket{1}_a\otimes\ket{1}_b=\ket{1}_b\otimes\ket{1}_a$ holds in this case.

To fully account for the CAR algebra we have to introduce the concept of braiding. This structure comes from the theory of quantum groups~\cite{braiding} and the main physical application today lies probably in the physics of anyons~\cite{anyons} and topological quantum field theory~\cite{TQFT}. A braided (or twisted) tensor product
\begin{equation}\label{eq:braidedtensor}
    \ket{1}_a\oxb\ket{1}_b=e^{i\vp}\ket{1}_b\oxb\ket{1}_a
\end{equation}
describes the whole family of statistics which, in a certain sense, interpolates between bosons and fermions. As a matter of fact, fermions are the simplest non-trivial example of the braided statistics where $\exp{i\vp}=-1$.
Hence, we have to understand Eq.~(\ref{eq:notensor}) in the following way:
\begin{equation}\label{eq:twistedtensor}
    a^\dg b^\dg\ket{vac}\equiv a^\dg\oxb b^\dg\ket{vac}=\ket{1}_a\oxb\ket{1}_b.
\end{equation}
To avoid clumsy notation in the rest of the text we will understand the braided tensor product for $\vp=\pi$ and abbreviate the RHS of Eq.~(\ref{eq:twistedtensor}) as $\ket{11}_{ab}$. But we will keep in mind that it is not an ordinary tensor product.

To formalize the findings we denote $\FF$ to be the Fock space of $k$ fermions. It is a $\bbZ_2$-graded vector space
\begin{equation}\label{eq:fermionFock}
    \FF=\FF^0\oplus\FF^1,
\end{equation}
where $i=\{0,1\}$ is the degree of $\FF^i$ corresponding to the even and odd Fock subspace. Following our discussion after Eq.~(\ref{eq:braidedtensor}), there exists a commutativity morphism $m$~\cite{braiding}
\begin{equation}\label{eq:morphism}
    m:\FF^i\otimes\FF^j\mapsto\FF^j\otimes\FF^i
\end{equation}
acting as
\begin{equation}\label{eq:morphism_acting}
    m(f^i\otimes f^j)=(-1)^{ij}f^j\otimes f^i
\end{equation}
on homogeneous elements $f^i\in\FF^i$. Note that $f^i$ labels both even and odd operators.

The adjoint operator for fermions is well defined and acts as
\begin{equation}\label{eq:daggeraction}
\dg:a^\dg\overset{\sim}{\otimes} b^\dg\mapsto -a\overset{\sim}{\otimes}b.
\end{equation}
Eq.~(\ref{eq:daggeraction}) necessarily follows from the anticommutation relations for particles obeying the fermion statistics and the normalization condition for fermion multi-particle  states:
\begin{equation}\label{eq:adjointactingonVAC}
    -\bra{vac}\big(a\oxb b\big)\big(a^\dg\oxb b^\dg\big)\ket{vac}
    =-\bra{11}_{ab}\ket{11}_{ab}=\bra{11}_{ba}\ket{11}_{ab}=1.
\end{equation}
One of the errors committed in~\cite{aunruh-fermi} was not to follow the above consistent rules. We
exemplify it on a fermion state which, as the authors of~\cite{aunruh-fermi} claim, is entangled and separable at the same time depending on the mode order. The state is actually not even allowed to exist according to the superselection rule for bosons and fermions but let us ignore this `detail' for the argument's sake. We will get to this point in a moment. The  disputed state reads
\begin{equation}\label{eq:threeparticlestate}
    \ket{\Phi}_{abc}={1\over2}(\ket{100}+\ket{010}+\ket{101}+\ket{011})_{abc},
\end{equation}
where we explicitly indicate the mode labels ($\ket{0}\equiv\ket{vac}$). As argued above, the Hermitian conjugation is a braided operator~\cite{braiding} and so for fermion fields it acts as
\begin{equation}\label{eq:threeparticlestate_bra}
    \dg:\ket{\Phi}_{abc}\mapsto\bra{\Phi}_{abc}={1\over2}(\bra{100}+\bra{010}-\bra{101}-\bra{011})_{abc}.
\end{equation}
We check the norm and find it positive $\brk{\Phi}{\Phi}=1$. Note that the adjoint operator acts on $\ket{\Phi}$ in the sense of Eq.~(\ref{eq:daggeraction}). To trace over a subsystem (for instance, the particle occupying the $c$-mode) we follow the rules of the CAR algebra and we obtain
\begin{equation}\label{eq:traceover}
    \Tr{c}{\kbr{\Phi}{\Phi}}={1\over2}(\ket{10}+\ket{01})_{ab}(\bra{10}+\bra{01})_{ab}.
\end{equation}
Following~\cite{aunruh-fermi}, we change the order of the modes $c$ and $b$ in~(\ref{eq:threeparticlestate}) and get
\begin{equation}\label{eq:threeparticlestate_swapped}
    \ket{\Phi'}_{acb}={1\over2}(\ket{100}+\ket{001}+\ket{110}-\ket{011})_{acb}.
\end{equation}
By tracing over the $c$ mode in $\kbr{\Phi'}{\Phi'}$ we have to again consistently use the CAR algebra. In this case, when we `skip' the occupied modes $a$ or $b$ by $\ket{1}_c$ or $\bra{1}_c$ the sign has to change. This follows from~Eq.~(\ref{eq:interCAR}). As a consequence, we again end up with state~(\ref{eq:traceover}) contrary to the conclusion from~\cite{aunruh-fermi}. A different result would be a sign of inconsistency of the whole CAR algebra.

However, there is even a more serious issue with the claims in~\cite{aunruh-fermi}. The states on which the authors demonstrate various inconsistencies violate the superselection rule for bosons and fermions. According to this rule the Fock space of all particles is a direct sum of a completely symmetric and antisymmetric subspace (for bosons and fermions, respectively, and in the usual 1+3 Minkowski spacetime). Therefore, one cannot form superpositions of fermions and bosons~\cite{QFT}. A closely related superselection rule forbids forming of superpositions of an even and odd number of fermions. If we take the antiunitary time reflection operator $T$ it is known~\cite{QFT} that
\begin{equation}\label{eq:timerefl}
    [T^2,a]_+=0
\end{equation}
holds for the spin one-half fermion annihilation operator $a$ (for the creation operator $a^\dagger$ as well). Clearly, the same transformation ensures that
\begin{equation}\label{eq:timerefl2}
    [T^2,ab]_-=0,
\end{equation}
where $[,]_-$ labels the commutator and $b$ is another fermion annihilation operator. Note that $T^2$ is a unitary operator and is not proportional to an identity. As a consequence, the Hilbert space is split into two sectors of even and odd fermion states~\cite{QFT} and no superpositions `connecting' the sectors can be formed. We now see why the state from Eq.~(\ref{eq:threeparticlestate}) is forbidden.

We can also resolve the second `paradox' encountered in~\cite{aunruh-fermi}. There, another and even simpler forbidden fermion state is considered
\begin{equation}\label{eq:Psi}
\ket{\Psi}_{ab}={1\over2}(\ket{00}+\ket{01}+\ket{10}+\ket{11})_{ab}.
\end{equation}

The authors argue~\cite{privatecom} that even after taking into account the right tensor product structure the state $\ket{\Psi}$ can be entangled and separable at the same time depending on the mode order. Based on the previous discussion, we might dismiss $\ket{\Psi}$ right away but let us proceed and see the consequences. We write
\begin{equation}\label{eq:Psi_abused}
    \ket{\Psi}_{ab}={1\over{2}}\Big(\ket{0}+\ket{1}\Big)_a\oxb\Big(\ket{0}+\ket{1}\Big)_b
    ={1\over2}(\id+a^\dg)\oxb(\id+b^\dg)\ket{0},
\end{equation}
where $\id$ stands for an identity operator. This is, however, a blatant violation of the superselection rule -- the expression $(\id+a^\dg)\ket{0}$ is not considered to be physical in QFT. If we continued this way we could have also written (still in accord with the rules of the CAR algebra)
\begin{equation}\label{eq:Psi_abusedagain}
    \ket{\Psi}_{ba}={1\over2}(\ket{00}+\ket{01}+\ket{10}-\ket{11})_{ba}.
\end{equation}
Indeed, if $\ket{0}$ and $\ket{1}$ carried the fundamental representation of the $SU(2)$ group, such state would be considered maximally entangled. This is not the case and we have a seemingly paradoxical situation where one state is factorizable and maximally entangled at the same time.

One could try to save the situation by assisting the states from Eqs.~(\ref{eq:Psi}) and (\ref{eq:Psi_abusedagain}) with two other modes, say $c$ and $d$, such that all kets would contain an equal number of fermions. The superselection rule is not violated and different partitioning (for example $ab$ and $cd$ versus $ad$ and $bc$) would indeed lead to a different amount of correlations. It is a well know fact that entanglement is partition-dependent. This is, however, not the case originally studied in~\cite{aunruh-fermi}. The states analysed here always violate the superselection rule and for those that are composed of two-fermions  no finer partitioning is possible.

We make a comment before we proceed. The difference between qubits  and fermions is that the  representation theory of the former is based on the $su(2)$ algebra while fermions are the field excitations generated by the fermion creation/annihilation operators obeying the CAR algebra. Consequently, fermions behave in a radically different way from qubits and when working with them one has to follow the rules of the CAR algebra with no exception. Qubits are quantum states living in an abstract Hilbert space spanned by a logical basis. By the term logical basis we simply mean that there are no qubit particles in Nature (cf.~\cite{stq,qubits}).
Any qubit prepared in a laboratory might either be  a spin one-half boson  or eventually a composite particle made of bosons or fermions. In the first case there  aren't that many candidates among elementary particles. Photons with the polarization encoding are qubits almost by an accident (photons are actually spin-one bosons but fortunately massless) and, in addition, the helicity subspace cannot be disentangled from the momentum degrees of freedom if relativistic transformations are taken into account~\cite{PT}.

In the second case, some effort must be invested to show that the composite particle carries a representation of the $SU(2)$ group. A similar statement holds for qudits in case of the $SU(d)$ group. But this is not happening in~\cite{aunruh-fermi}. The mapping used by the authors is clearly not a representation of $SU(2)$. Their construction~\cite{privatecom} consists of substituting the fermion operators
by the operators satisfying the canonical commutation relations (boson creation/annihilation operators). They repeat the same procedure for a different order of fermions and they treat the resulting states as different qubit multi-partite states. As two-qubit states they would have been indeed different and there is no need why, for example, the procedure of partial trace in~Eqs.~(\ref{eq:Psi}) and~(\ref{eq:Psi_abusedagain}) would lead to the same result. But let us make it clear:
\begin{itemize}
  \item The procedure of substituting fermions for qubits as done in~\cite{aunruh-fermi} is flawed -- fermions do not become qubits just by changing their name.
\end{itemize}
The objects living in $\FF^1$ in Eq.~(\ref{eq:fermionFock}) have to strictly follow the rules fermions obey. Odd fermion operators cannot be substituted by even fermion operators. The tensor product structure is crucial for avoiding contradictions \`a la~\cite{aunruh-fermi}. This is the main source of confusion in~\cite{aunruh-fermi} but not in~\cite{fermi1a,fermi2} contrary to the claim of the authors.

It would have been a completely different story if they substituted fermions from the even Fock subspace $\FF^0$. An even number of fermions commute and such objects behave as distinguishable bosons in the sense of Eq.~(\ref{eq:morphism_acting}) (set $i=j=0$)~\footnote{Note that distinguishable bosons can form any superposition they wish. The symmetrization procedure does not apply. One can prepare the singlet state $\ket{\Psi}={1/\sqrt{2}}(\ket{01}-\ket{10})_{ab}$ out of two bosons $a$ and $b$ with two orthogonal polarization degrees of freedom denoted by $0$ and $1$.}. Clearly, this kind of composite particles are not true bosons since they don't satisfy the canonical commutation relations (cf.~\cite{law}). But this is not the goal one aims for anyway. It is the tensor product structure of such objects that matters so if there is a chance to consider fermions as qubits it only lies in the even subspace of the fermion Fock space. We will stop here and point to one such construction in the context of Unruh effect presented in~\cite{grass}. The braiding properties of fermions and the superselection rule have fully been considered and for a carefully chosen encoding one can indeed form qubits out of fermions lying in the even sector of the Fock space $\FF$. Certain entropic quantities used in quantum Shannon theory were proved to remain valid but only for a restricted class of fermion states very different from the ones used in~\cite{fermi1,fermi1a,fermi2}. 

In passing, let us present two examples of fermion states from the even sector $\FF^0$ of the total Hilbert space. One such example is
$$
\ket{\Omega}_{ab}={1\over\sqrt{2}}(\ket{vac}+\ket{1}_a\ket{1}_b)
\equiv{1\over\sqrt{2}}(\ket{0}_a\ket{0}_b+\ket{1}_a\ket{1}_b),
$$
where we slightly abuse notation in the rightmost equation. Another example based on two distinguishable spin one-half fermions (so-called Dirac particles) reads
$$
\ket{\Omega}_{ab}={1\over\sqrt{2}}(\ket{\uparrow}_a\ket{\downarrow}_b+\ket{\downarrow}_a\ket{\uparrow}_b),
$$
where $\ket{\uparrow(\downarrow)}_a=\ket{1}_{a,\uparrow(\downarrow)}=a_{\uparrow(\downarrow)}^\dg\ket{vac}$ (only at this occasion the spin degree of freedom is explicitly indicated by an arrow and the momentum degree of freedom is the letter itself). Unless other superselection rules become relevant such states can be formed. These states are not analyzed in~\cite{aunruh-fermi}.

To summarize, the authors of~\cite{aunruh-fermi} cannot expect to get consistent results when deciding whether a fermion state is separable or not if the object they deal with is profoundly ill-defined and, moreover, the basic rules of the CAR algebra are not being respected. But the purpose of this note is not to criticize a single wrong article. It is the actual motivation for~\cite{aunruh-fermi} that feels uneasy. The principal observation there is that many results on fermion quantum correlations in the context of Unruh effect~\cite{fermi1,fermi1a,fermi2} are not compatible. But as we said at the beginning, a quick glance over~\cite{fermi1a,fermi2} suggests that the authors did not fall into the same QFT trap as in~\cite{aunruh-fermi,fermi1}. The issue is more subtle and lies in one of the consequences of the superselection rule:
\begin{itemize}
  \item Criterions for deciding whether a fermion multi-particle  state is separable or how much quantum correlations it contains cannot be adopted from the qubit case without further reasoning.
\end{itemize}
This fact has already been explicitly shown by the authors of~\cite{braiding_sep} and they found that the only allowed two-fermion separable state is
$$
\vr_{sep}=\diag{[\nu_1,\nu_2,\nu_3,\nu_4]},
$$
where $0\leq\nu_i\leq1,\sum\nu_i=1$. Therefore, a two-qubit state that would be classified as separable might not be written in a separable form if it was a two-fermion state. It would violate the superselection rule.

Clearly, the same conclusion holds for entanglement measures. Hence, even if have shown that the calculations of various entropic quantities (for example, the  often calculated (logarithmic) negativity~\cite{neg} in~\cite{aunruh-fermi,fermi1,fermi1a,fermi2}) can be made consistent even for fermions, we do not claim that these  quantities  have actually the same meaning or can be interpreted in the same way as for qubits. This is the most likely culprit for the different results found in literature~\cite{fermi2}. As an example, let us take a closer look at~\cite{fermi1a} which is one of the first articles studying entanglement in mixed states of two fermions. The authors work with full-fledged fermions (contrary to the claim in~\cite{privatecom}) and they are clearly aware of the subtleties coming from QFT. But it is in the quantum information section where the problem resides.

To see where exactly the problem is we first notice that no fundamental principle prevents us from adopting the entanglement of formation as the unique definition of entanglement~\cite{EoF} for bipartite fermion systems
\begin{equation}\label{eq:EoF}
    E_F(\vr_{ab})=\min\sum_ip_iH(\vr_a^i),
\end{equation}
where $H(\sigma)$ stands for the von Neumann entropy of $\sigma$ and $\sum_ip_i=1$. The minimum is taken over all possible decompositions of $\vr_{ab}$. But at this point the superselection rule enters the game. Some decompositions acceptable for a two-qubit system are not allowed for two fermions. So we cannot use the Wootters concurrence~\cite{Woo} to calculate the entanglement of formation as we would have done it for qubits. The qubit case is at best a lower bound and we will see that not very tight. The authors of~\cite{braiding_sep} went further and actually found an explicit formula for the entanglement of formation of a general two-fermion mixed state. We can therefore compare the entanglement calculations, for example, for state (22) in~\cite{fermi1a}
\begin{equation}\label{eq:eq22}
    \vr_{ab}={1\over2}\begin{pmatrix}
      \cos{r}^2 & 0 & 0 & \cos{r} \\
      0 & \sin{r}^2 & 0 & 0 \\
      0 & 0 & 0 & 0 \\
      \cos{r} & 0 & 0 & 1 \\
    \end{pmatrix}.
\end{equation}
\begin{figure}[t]
\begin{center}
    \resizebox{13.5cm}{9cm}{\includegraphics{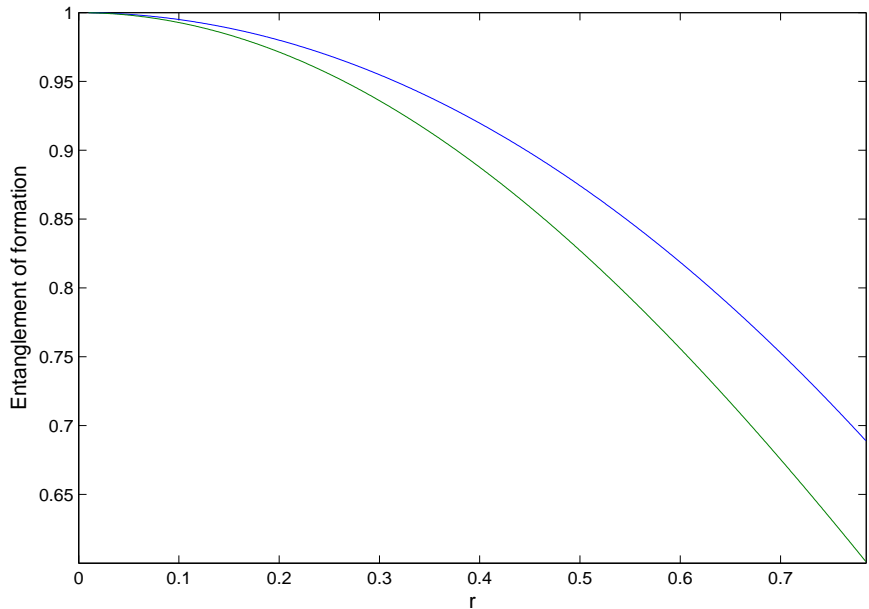}}
    \caption{The entanglement of formation for a fermion mixed state in Eq.~(\ref{eq:eq22}) taken from~\cite{fermi1a} (Eq.~(22)). The lower curve is the original calculation. The upper curve is the correct calculation based on~\cite{braiding_sep} taking into account the superselection rule for fermions.}
    \label{fig:eof}
\end{center}
\end{figure}
The state is written in the fermion basis $\{\ket{00},\ket{01},\ket{10},\ket{11}\}_{ab}$ and $r$ is the acceleration parameter ($r=0$ corresponds to no acceleration and $r=\pi/4$ is the infinite acceleration limit). The result is depicted in Fig.~\ref{fig:eof} and with the increasing acceleration the gap between the two calculations widens.

No need to say that the situation is even worse for tripartite correlations. The main tool to investigate them is the negativity as well. For tripartite qubit states the negativity is already a mere bound since it does not account for non-distillable entanglement. These two problems together indicate that the calculations of tripartite correlations in~\cite{fermi1a,fermi2} are approximate at best.

\section{Entanglement fallacy}

The second part of this text is about the motivation behind the various calculations trying to quantify quantum and classical correlations of states undergoing Unruh or Hawking transformation. The field of relativistic quantum information studies the behavior and flow of quantum information in a relativistic setting. If we disregard the relativistic part for a while there is a large subdiscipline of quantum information science called quantum Shannon theory~\cite{shannon}. Quantum Shannon theory generalizes classical Shannon theory~\cite{coverthomas} to the quantum domain. It has been extremely successful in describing and quantifying the flow of quantum and classical information in quantum systems. The main goal of the theory is to find capacity formulas which are designed  to capture how much information can be reliably transferred in a quantum system. There exist several definitions of capacities depending on what type of information one is interested (classical versus quantum) or what type of resources are available.

The methods of quantum Shannon theory are perfectly suited for the realm of relativistic quantum information~\cite{grass,unruh-bosonic,vacchannel}. The reason is the relativistic setup itself. Quantum-relativistic effects usually occur under somewhat extreme conditions. It may be high velocities, huge masses, large distances, great emptiness or all together. The large-scale structure of spacetime comes into play and it is very natural to ask how much information can be transmitted from one spacetime point to another and under what conditions. The answers are provided by quantum Shannon theory. The capacity of a quantum channel is a fundamental notion telling how much information {\em in principle} can be extracted or transmitted through a quantum system. the reason is simple. No seriously meant quantum communication scenario (private or not) will use a single copy of some quantum resource. Quantum Shannon theory is an asymptotic theory and the mighty machinery of quantum error-correction is available. As a result, one can claim whether classical or quantum communication over a quantum channel is principally possible. As an example we offer the teleportation protocol. It is difficult to distribute maximally entangled states at large distances and no one would even bother to do so just to transfer a few qubits. Instead, the quantum channel is identified between a sender and receiver and if its quantum capacity is nonzero we know that in an asymptotic limit of many copies of the channel the result of quantum error-correction is the creation of a new channel (effectively an identity channel). This channel enables a near-perfect transfer of many maximally entangled states. Quantum communication is therefore possible. It is exactly this scenario that saves the situation for a teleportation to a non-inertial frame. While originally concluded to be impossible~\cite{noninertialtele}, it can be shown that reliable quantum transmission is possible for any finite acceleration~\cite{unruh-bosonic,classcomm}.

Note that the asymptotic character of quantum Shannon theory is a theoretical assumption. In practice, if a suitable quantum code is found,  the number of quantum states used to assemble the  code is relatively reasonable and it approaches the theoretical Shannon limit fast enough.

Instead of adopting the information-theoretic approach, the focus of a majority of articles in relativistic quantum information lies in the investigation and quantifying the entanglement or other quantum correlations of a relativistically transformed state. The amount of entanglement is certainly an important quantity that gives some information about the behavior of a quantum system~\cite{shi}. It might be interesting to know how quantum correlations behave under relativistic transformations for its own sake. It also may have an impact on our understanding of the role of entanglement on a cosmological scale~\cite{meni} and  other conceptual issues~\cite{conceptual}. However, the  notion of entanglement for the purpose of quantum information transmission is desperately uninformative or even misleading. We will call it the entanglement fallacy. Problems like the quantum communication of two relativistic observers or the black hole information paradox are exactly the type of situations where quantum Shannon theory and not entanglement measures or similar simple concepts will be able to provide the ultimate answer or explanation.

We will present the raised issue on two more modest problems. First we will bring an example of a quantum channel whose parameter can be tuned such that its capacity goes from the maximum one bit to zero yet the entanglement is never destroyed. This example will show that even if the entanglement measures were calculated correctly for fermion systems in non-inertial frames and the problems of the first section of this text were avoided we should still be careful when interpreting those results. The channel is a qubit erasure channel~\cite{erasureCh} $\N:\B(\bbC^2)\mapsto\B(\bbC^3)$ acting as
$$
\N(\vr)=(1-p)\vr+p\kbr{e}{e}.
$$
The state $\ket{e}$ is a flag state such that $\brk{e}{0}=\brk{e}{1}=0$ holds. The Kraus operators take the following form:
\begin{subequations}\label{eq:erasureChKraus}
\begin{align}
    N_1&=\sqrt{1-p}\begin{pmatrix}
          1 & 0 \\
          0 & 1 \\
          0 & 0 \\
        \end{pmatrix}\\
    N_2&=\sqrt{p}\begin{pmatrix}
          0 & 0 \\
          0 & 0 \\
          0 & 1 \\
        \end{pmatrix}\\
    N_3&=\sqrt{p}\begin{pmatrix}
          0 & 0 \\
          0 & 0 \\
          1 & 0 \\
        \end{pmatrix},
\end{align}
\end{subequations}
where $0\leq p\leq1$. The channel plays quite an important role in quantum information theory~\cite{erasureCh}. An interesting coincidence is that a qubit erasure channel also appears in the context of the Unruh effect for fermions~\cite{grass} and we will get to this point later.

It is known that the quantum capacity is non-zero for $0\leq p<1/2$. What happens if we send a half of a maximally entangled state $\Psi$  through the channel? We find the eigenvalues for the partial transpose of the state $\Phi=(\id\otimes\N)(\Psi)$. The eigenvalues are
\begin{equation}\label{eq:PPT}
\rm{Eig}[\Phi^\Gamma]=\left[{1\over2}(p-1),{1\over2}(1-p),{1\over2}(1-p),{1\over2}(1-p), {p\over2},{p\over2}\right]
\end{equation}
and the PPT criterion tells us that the state $\Phi$ is separable only for $p=1$. Hence for the whole interval $p\in[0,1)$ the erasure channel produces (partially) entangled states but this is useless for quantum communication for $p\in[1/2,1)$. We therefore cannot take the existence of entanglement as a trusty indicator that the physical system described by a qubit erasure channel can actually reliably convey quantum information.

A qubit erasure channel is more relevant for the propagation of quantum information to non-inertial frames than it seems. In~\cite{grass} we have shown that this channel is precisely the qubit Grassmann channel which is the embodiment of the
the Unruh effect for fermions. The structure and construction of the Grassmann channel fully respects the rules of the CAR algebra and the superselection rule to avoid the troubles from the first section.

In the literature on the fermion Unruh effect~\cite{fermi1a,fermi2} a lot of attention is paid to the differences in the entanglement behavior between Unruh effect for bosons and fermions. Especially the infinite acceleration limit is often calculated and the asymptotic analysis predicts that for bosons the entanglement dies out~\cite{unruh-bosons} while for fermions not~\cite{fermi1a,fermi2}. As we see in Fig.~\ref{fig:eof} this result is likely to survive the serious objections raised in the first section. There is, however, no use for such entanglement from the quantum communication point of view. We calculated the quantum capacity of the qudit Grassmann channel for an arbitrary $d$~\cite{grass} and compared it with the boson case~\cite{unruh-bosonic}. The capacities behave in the same way for both cases. It monotonically decreases with an increasing acceleration and in the infinite acceleration limit the quantum capacity approaches zero at the same rate for bosons and fermions. For qubits ($d=2$), the value for which the capacity is zero corresponds  to the qubit erasure channel for $p=1/2$ in Eqs.~(\ref{eq:erasureChKraus}) and from Eq.~(\ref{eq:PPT}) we see that in this case $\Phi$ is entangled.
Therefore, there is no qualitative difference between the Unruh effect for bosons and fermions and the entanglement of shared fermion states in the limit is useless. It cannot be utilized to reliably transmit quantum information. The implication actually goes  in the opposite way: If the quantum capacity is non-zero a near-perfect entanglement transfer is possible. This is one of the operational interpretations of the quantum capacity. The quantum capacity informs us how big portion of an input Hilbert space can be coherently transmitted between a sender and receiver or the ratio of initially maximally entangled states that will `pass through' a quantum channel undisturbed.

\section{Conclusions}

In this note we pointed out two misconceptions that plague current research in relativistic quantum information. The first issue is related to studies of the Unruh effect for fermions where entanglement behavior is investigated. Various investigators found different answers to similar questions regarding the behavior of quantum correlations in non-inertial frames. We identified several reasons why there is so many incompatible results. The basic error is the improper use of the fermion algebra and ignorance of the superselection rule for bosons and fermions as is customary in quantum field theory. This consequently leads to a number of illusory `paradoxes'. This type of error, however, seems to be avoided by the majority of articles dealing with this topic. The main problem lies in one of the consequences of the superselection rule: the misinterpretation of entanglement measures (and quantum correlation measures in general) for multi-partite qubit states. The same measures are used to evaluate quantum correlations of fermion multi-particle states. But this `transfer' cannot be done without taking into account the superselection rule for bosons and fermions.

The second discussed point in this note is not a calculational mistake but rather an interpretation issue. We argue that for relativistic scenarios treating quantum information in non-inertial frames it is the  transmission of quantum information that is important. Therefore, there is not much sense in calculating entanglement measures since, as we demonstrated, they don't really reliably inform whether  reliable quantum information transfer is possible. This is precisely the piece of information that quantum channel capacities provide as investigated by quantum Shannon theory.

\acknowledgements
The author would like to thank Prakash Panangaden, Roc\'io J\'auregui and Tomas Jochym-O'Connor for discussions. This work was supported by a grant from the Office of Naval Research (N000140811249) and some of its parts were conceived during the NITheP Workshop on Relativistic Quantum Information organized by the Centre for Quantum Technology, University of KwaZulu-Natal.

\end{document}